\documentclass[aip,reprint,amsfonts,amsmath,amssymb,superscriptaddress,10pt,floatfix]{revtex4-1}

\usepackage[english]{babel}
\usepackage[utf8]{inputenc}

\usepackage{amsmath}
\usepackage{amsfonts}
\usepackage{amssymb}
\usepackage{amsthm}
\usepackage{mathtools}
\usepackage{graphicx}
\usepackage{xfrac}

\usepackage{color}
\usepackage{array}
\usepackage{longtable}
\usepackage{multirow}

\begin{document}

\title{A string reaction coordinate for the folding of a polymer chain}

\author{Christian Leitold}
\affiliation{Faculty of Physics, University of Vienna, Boltzmanngasse 5, 1090 Vienna, Austria}

\author{Wolfgang Lechner}
\affiliation{Institute for Quantum Optics and Quantum Information, Austrian Academy of
Sciences, Technikerstraße 21a, 6020 Innsbruck, Austria}
\affiliation{Institute for Theoretical Physics,  University of Innsbruck, Technikerstraße 21a, 6020 Innsbruck, Austria}

\author{Christoph Dellago}
\affiliation{Faculty of Physics, University of Vienna, Boltzmanngasse 5, 1090 Vienna, Austria}

\date{December 4, 2014}

\begin{abstract}
We investigate the crystallization mechanism of a single, flexible homopolymer chain with short range attractions. For a sufficiently narrow attractive well, the system undergoes a first-order like freezing transition from an expanded disordered coil to a compact crystalline state. Based on a maximum likelihood analysis of committor values computed for configurations obtained by Wang--Landau sampling, we construct a non-linear string reaction coordinate for the coil-to-crystal transition. In contrast to a linear reaction coordinate, the string reaction coordinate captures the effect of different degrees of freedom controlling different stages of the transition. Our analysis indicates that a combination of the energy and the global crystallinity parameter $Q_6$ provide the most accurate measure for the progress of the transition. While the crystallinity paramter $Q_6$ is most relevant in the initial stages of the crystallization, the later stages are dominated by a decrease in the potential energy.
\end{abstract}

\maketitle

\section{Introduction}

Many polymers go through large-scale conformational changes akin to phase transitions when external parameters like the temperature or the solvent properties change~\cite{degennes_scaling}. A particularly simple example for such a system is a homopolymer chain with short-range attractions and strongly repulsive cores~\cite{tpb_1}. The complex phase behavior of this system has been studied previously in a number of different studies. First, it was shown by Taylor and Lipson~\cite{taylor_ex1, taylor_ex2}, as well as later by Zhou~et\,al.~\cite{zhou}, that the chain's radius of gyration is a sigmoidal function of temperature. More recently, the entire phase diagram of the system as a function of temperature and interaction range was mapped out by Taylor, Paul, and Binder~\cite{tpb_1, tpb_2, tpb_4}. Of particular interest to this study is the first-order like coil-to-crystal freezing transition, which occurs for chains with very narrow attractive wells. This transition has been studied recently by Růžička, Quigley, and Allen~\cite{allen_folding} in forward flux sampling simulations of a slightly modified model to allow the application of collision dynamics. In a further study, the authors of this paper have investigated the freezing transition using transition path sampling~\cite{tps0} in combination with likelihood maximization~\cite{peters_rc} in order to search for a reaction coordinate~\cite{Leitold2014}. Here, we build on this work and improve the quality of the reaction coordinate by substituting the linear version used so far with a non-linear string reaction coordinate~\cite{Lechner2010}.

The remainder of this paper is organized as follows. In Sec.\,\ref{sec:model} we define the polymer model and give a short summary of its properties. Methods and simulation details are discussed in Sec.\,\ref{sec:methods}. We present results in Sec.\,\ref{sec:results}, and provide a discussion in Sec.\,\ref{sec:discussion}.

\section{Polymer model}
\label{sec:model}

\begin{figure}
\includegraphics[width=0.95\columnwidth]{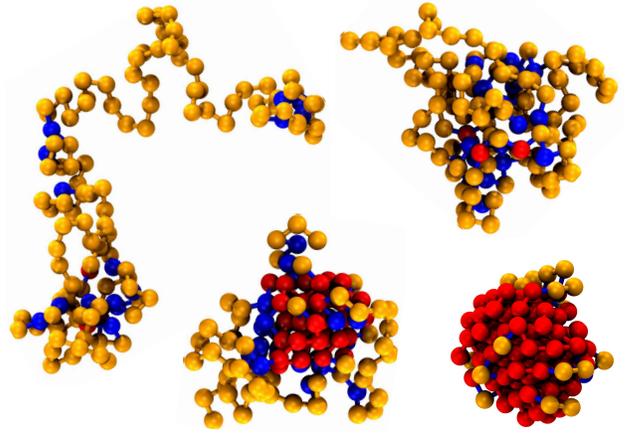}
\caption{Coil (top left), crystalline (bottom right), and two intermediate states of the polymer chain for particle number ${N=128}$, interaction range ${\lambda = 1.05}$ and temperature ${k_{\rm{B}} T / \varepsilon =0.438}$. Crystalline and coil-like particles are colored in red and yellow, respectively, while intermediate particles are colored in blue. The criterion for crystallinity used here is defined in Ref.\,\onlinecite{Leitold2014}.}
\label{fig:states}
\end{figure}

\begin{figure}
\includegraphics[width=0.95\columnwidth]{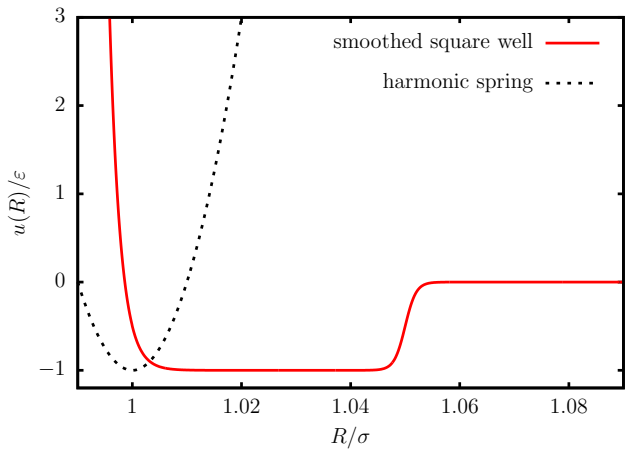}
\caption{Smoothed square-well potential for ${\lambda = 1.05}$. The harmonic spring potential (acting between neighboring beads only) is also shown for comparison.}
\label{fig:potentials}
\end{figure}

The model used in this study is a single, fully flexible chain of $N$ identical monomers with a short-range attraction between monomers, as well as a hard repulsive core (Fig.\,\ref{fig:states}). Non-neighboring monomers interact via a smoothed variant of a square-well potential~\cite{Leitold2014},
\begin{equation}
u(R) = \frac{\varepsilon}{2} \left\{ \exp \left[ \frac{-(R-\sigma)}{a} \right] + \tanh \left[ \frac{R-\lambda \sigma}{a} \right] - 1 \right\},
\label{eq:potential}
\end{equation}
where $R$ is the distance between the monomers and ${\lambda > 1}$ parametrizes the width of the potential well. We have chosen a value of ${a = 0.002 \, \sigma}$ for the parameter which determines the steepness of the exponential repulsion and the width of the step at ${R = \lambda \sigma}$. Neighboring monomers are coupled via harmonic springs ${U(R) = \frac{k}{2} (R-\sigma)^2}$ with a value of ${k= 20000 \, \sigma^2 / \varepsilon}$ for the spring constant. The pair potential, as well as the harmonic potential between chain neighbors, is shown in Fig.\,\ref{fig:potentials}.

Depending on the value of the interaction width $\lambda$, there exist three different phases. At high temperatures, the system is in the expanded coil phase for all interaction widths. What happens when the system is cooled, however, depends on $\lambda$. For wide wells, the system first undergoes a second-order collapsing transition to a compact, but unordered globule phase. Further cooling leads to a first-order freezing transition to a crystalline state. For sufficiently small values of $\lambda$ (${\lambda \leq 1.05}$ in the case of ${N=128}$), the system directly freezes from the coil to the crystalline state without going through the molten globule phase. A detailed description on the chain's phase behavior is given in the work of Taylor, Paul and Binder~\citep{tpb_1, tpb_2, tpb_4}, as well as our own recent study~\cite{Leitold2014}. In the latter work, we also show that the phase behavior of the smoothed version of the chain is very similar to that of the original square-well chain. For simulations presented and discussed in this paper, we have chosen the ${N=128}$ chain with an interaction length of ${\lambda = 1.05}$. This system undergoes a direct freezing transition from the coil to the crystalline state, with a coexistence temperature of ${k_{\rm{B}} T / \varepsilon = 0.438 \pm 0.001}$~\cite{Leitold2014}. An illustration of these two states is shown in Fig.\,\ref{fig:states}.

\section{Methods}
\label{sec:methods}

\subsection{Definition of the stable states}

In this paper we study the freezing of the polymer. In accordance with our previous study of this system~\citep{Leitold2014}, we define the expanded coil as stable state $A$ and the crystal as stable state $B$. The distinction is made based on the potential energy of the system. A configuration is considered to be in the coil state if ${U/N \geq U_{\rm{min}}/N = -0.7 \, \varepsilon}$ and it is considered to be in the crystalline state if ${U/N \leq U_{\rm{max}}/N = -2.6 \, \varepsilon}$. Note that in our model the potential energy is essentially proportional to the number of close contacts between non-neighboring monomers.

\subsection{Wang--Landau sampling}

Using a Wang--Landau simulation~\cite{wang-landau} we have obtained a uniform sample of states outside the two stable basins $A$ and $B$. With this technique, one iteratively constructs the density of states of the system by performing a Monte Carlo simulation with the inverse of the current estimate of the density of states as acceptance criterion. Once the simulation is converged, each energy interval of a given fixed width is visited with equal frequency. To speed up the simulation we have combined several types of Monte Carlo Moves including the bond-bridging move introduced in Ref.\,\onlinecite{tpb_2}. Further details of the Wang--Landau procedure, as well as the Monte Carlo moves used in the simulation, are given in Ref.\,\onlinecite{Leitold2014}.

\subsection{Committor analysis}

For a system with two (meta-) stable states $A$ and $B$, the committor $p_B(\mathbf{x})$ of configuration $\mathbf{x}$ is the fraction of dynamical pathways started from $\mathbf{x}$ that first reaches state $B$~\cite{tps0}. To compute the committor for a particular $\mathbf{x}$, one launches a number of trajectories starting with random momenta from $\mathbf{x}$ and counts the fraction of trajectories ending in $B$. Our committor calculations were performed according to the algorithm described in Ref.\,\onlinecite{tps_fluid}, using ${N_{\rm{min}} = 100}$ and ${N_{\rm{max}} = 500}$. Since we are interested in the true mechanism of the coil-to-crystal transition, the procedure used to obtain trajectories for the calculation of committor values must resemble the natural dynamics of the system. If Monte Carlo dynamics is considered, this implies that only local moves can be used. Molecular dynamics provides a more physical (and computationally more efficient) way to model the time evolution of the system. We therefore use the smooth (differentiable) potential of Eqn.\,(\ref{eq:potential}) to facilitate such simulations and avoid the cumbersome handling of impulsive forces caused by the discontinuities in the original square-well potential. To evolve the system in time, we employ Langevin dynamics with a time step $\Delta t = 0.0002 \, \sqrt{m \sigma^2 / \varepsilon}$ and a damping constant $\gamma = 0.5 \, m^{3/2} \sigma^2 \varepsilon^{-1/2}$, where the actual integration is performed using the Langevin thermostat by Schneider and Stoll~\cite{schneider} implemented in a modified version of LAMMPS~\cite{lammps}.

\subsection{String reaction coordinate}

\begin{figure}
\includegraphics[width=0.7\columnwidth]{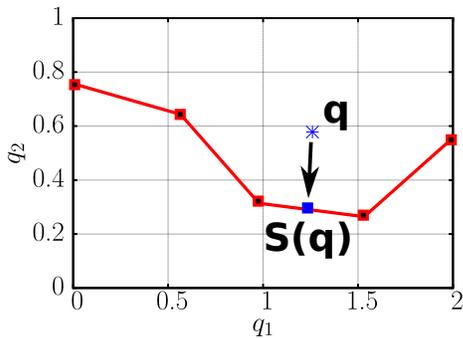}
\caption{Schematic representation (data not from simulation) of $\mathbf{S}^M(\mathbf{q})$, for a string of length ${M=5}$. The point $\mathbf{q}$ is first mapped to a point on the string $\mathbf{S}(\mathbf{q})$. This point is then mapped by $\alpha(\mathbf{S})$ to a number between $0$ and $1$ to obtain the progress along the string.}
\label{fig:S_of_q}
\end{figure}

To identify a valid reaction coordinate, we use the non-linear reaction coordinate analysis of Lechner {\em et al.}~\cite{Lechner2010} In this approach, a reaction coordinate is constructed as a projection of a configuration on a piecewise linear string defined by a sequence of $M$ points in an $m$-dimensional order parameter space. More formally, the reaction coordinate $r$ associated with a point $\mathbf{x}$ in configuration space is defined as
\begin{equation}
r(\mathbf{x}) = f\big(\alpha\{\mathbf{S}^M[\mathbf{q}(\mathbf{x})]\}\big).
\end{equation}
Here, we have a sequence of projection operations. $\mathbf{q}(\mathbf{x})$ maps the $3N$-dimensional configuration to a low-dimensional order parameter space, so ${\mathbf{q} = \{ q_1, \dots, q_m \}}$ is a vector of order parameter values. In this work, we restrict ourselves to ${m=2}$, in other words, the string resides in a two-dimensional plane. $\mathbf{S}^M(\mathbf{q})$ is the projection onto the string, schematically illustrated in Fig.\,\ref{fig:S_of_q} for a plane of two generic order parameter $q_1$ and $q_2$, and $\alpha(\mathbf{S}^M)$ is the mapping of the string point to a number between $0$ (state $A$, start of the string) and $1$ (state $B$, end of the string). We use the geometric projection described in the work of Rogal~et.\,al~\cite{Rogal2010}. Finally, $f(\alpha)$ is a monotonic (cubic) spline that maps $\alpha$ to the reaction coordinate.

The committor $p_B$ is assumed to be a sigmoidal function of the model reaction coordinate,
\begin{equation}
p_B(r) = \frac{1}{2} [1+\tanh(r)].
\label{eq:pbofr}
\end{equation}
The reaction coordinate, defined by the location of the string points, the relative scaling of the involved order parameters and the functional form of the projection, is constructed such that the likelihood
\begin{equation}
L = \prod_{k}^{B} p_B(r^{(k)}) \prod_{k'}^{A} [1-p_B(r^{(k')})]
\label{eq:Lalpha}
\end{equation}
is maximized. The first product runs over all single shooting events ending in state $B$, while the second product runs over all shooting events ending in state $A$. The likelihood quantifies the compatibility of the proposed model with observed outcome of the shooting events. We use the Bayesian information criterion~\cite{bic}
\begin{equation}
\mathrm{BIC} = {-2 \ln{L} + k(M) \ln(n)}
\end{equation}
to compare the optimization results for different numbers of optimization parameters, where smaller BIC values are better. Here, $n$ is the total number of observations, i.\,e., the total number of shooting events entering in Eq.\,(\ref{eq:Lalpha}), and $k(M)$ is the number of free parameters entering the model. The BIC penalizes models with too many free parameters, hence it is used to check whether it is sensible to add additional physical parameters to improve the model reaction coordinate. The first coordinate, in our study the polymer's potential energy, is used to distinguish the two stable states $A$ and $B$, therefore, the end points of the string are held fixed at the stable state boundaries and can only move in the orthogonal direction. The inner points of the string can move in any direction. In addition, we use $M$ equally spaced points between $0$ and $1$ to define the mapping $f(\alpha)$, as well as an additional variable for the scaling of the plane along the second variable relative to the first one. Therefore, ${k(M) = 3M - 1}$ for ${M \geq 2}$.

The actual optimization of the string is carried out according to the algorithm described in Ref.\,\onlinecite{Lechner2011a}. The procedure is a steepest descent scheme, where one move is either one of three different choices:
\begin{enumerate}
\item a string move, where the string itself is altered by displacing the points of the string;
\item a move where the mapping $f(\alpha)$, which translates the progress along the string to a reaction coordinate, is altered;
\item a scaling move, where the weight of the second variable relative to the first (the energy) is changed.
\end{enumerate}
In all cases, the move is accepted if the (log) likelihood increases and rejected otherwise.

\subsection{Order parameters}
\label{subsec:orderparameters}

In addition to the potential energy, we have calculated a number of structural order parameters for the polymer. These are
\begin{itemize}
\item the size of the core $N_{\rm{core}}$: the number of particles in the largest cluster of crystalline particles;
\item the total number of crystalline particles $N_{\rm{cryst}}$;
\item the total number of compact particles $N_{\rm{comp}}$, defined as all particles with six or more adjacent particles (distance smaller than $1.05 \, \sigma$);
\item the mean squared radius of gyration $R_g^2$;
\item the global order parameters $Q_4$ and $Q_6$;
\item the polymer's moments of inertia ${I_1 = I_{\min}}$, $I_2$, and ${I_3 = I_{\max}}$;
\item the anisotropy ${a = I_{\max} / I_{\min} - 1}$.
\end{itemize}
For all these variables, we have constructed string reaction coordinates with ${2 \leq M \leq 9}$ in combination with the potential energy $U$ as first coordinate. The criterion for crystallinity used in the calculation of $N_{\rm{cryst}}$ is defined in Ref.\,\onlinecite{Leitold2014}.

\section{Results}
\label{sec:results}

\begin{figure}
\includegraphics[width=1.0\columnwidth]{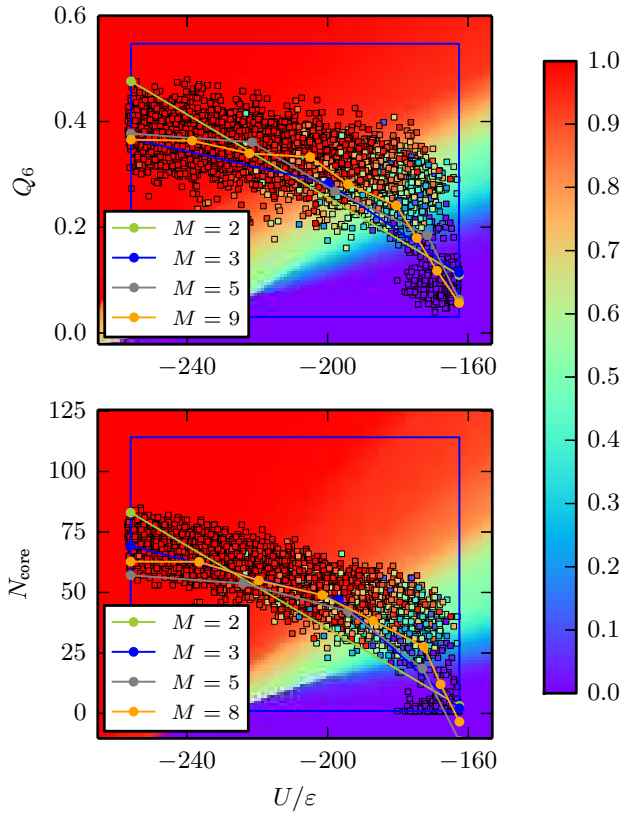}
\caption{(Top) Optimized strings with ${M=2, 3, 5, 9}$ in the $U$-$Q_6$ plane. The blue rectangle shows the range of order parameter values of the polymer configurations considered in the procedure. The color map in the background represents the predicted committor from the ${M=5}$ string. The colored dots are the real committor values of the configurations used in the optimization, with the same color code as the one used for the predicted values. (Bottom) Optimized strings with ${M=2, 3, 5, 8}$ in the $U$-$N_{\rm{core}}$ plane. The color map in the background represents the predicted committor from the ${M=8}$ string.}
\label{fig:string_image}
\end{figure}

\begin{figure}
\includegraphics[width=1.0\columnwidth]{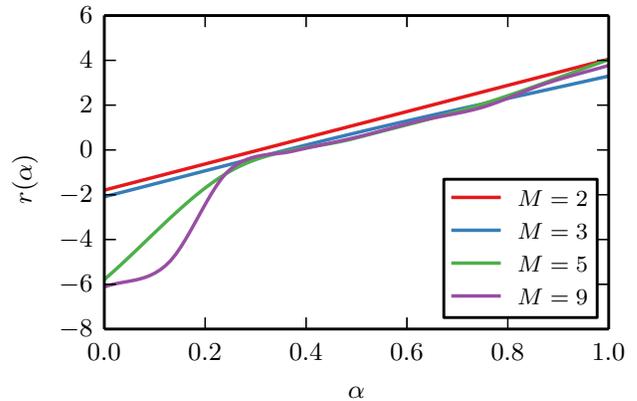}
\caption{Cubic spline mapping functions ${r = f(\alpha)}$ belonging to the four strings in the top panel of Fig.\,\ref{fig:string_image}.}
\label{fig:ralpha}
\end{figure}

We have used a total of 3912~configurations in the energy range ${-256 < U / \varepsilon < -163}$ with known committor values, corresponding to the left and right borders of the blue rectangles in Fig.\,\ref{fig:string_image}, for the construction of the string reaction coordinate. The string with ${M=2}$ corresponds to the linear model reaction coordinate as introduced by Peters and Trout~\citep{peters_rc}, so as a first test, we have checked that the results obtained for the ${M=2}$ string are in agreement with the results of our implementation of the linear optimization procedure. Apart from some discretization error due to the way the committor data are handled in our string coordinate code, these two likelihood scores agree well. The score for the linear version also serves as a baseline to compare the likelihood score for more complex strings. In particular, for any combination of variables, the likelihood score obtained by a string with ${M \geq 3}$ should be greater than the corresponding likelihood for the optimized linear reaction coordinate. It is worth noting that the main computational effort in the construction of the reaction coordinate goes into the calculation of committor values and has to be performed on a cluster with many computing cores. In contrast, the string optimization procedure, even for many combinations of variables, can be done at comparatively low computational cost on a single workstation.

\renewcommand*\arraystretch{1.2}
\renewcommand{\tabcolsep}{1.0em}
\begin{table}[htbp]

\begin{tabular}{l|r r r}
\hline \hline
 & \multicolumn{1}{l}{\rm{BIC}$_{\rm{min}}$} & \multicolumn{1}{l}{$\ln L$} & \multicolumn{1}{l}{$M$}\\
\hline
$Q_6$ & 231632 & -115725 & 5 \\
$N_{\rm{core}}$ & 237303 & -118502 & 8 \\
$N_{\rm{cryst}}$ & 237720 & -118711 & 8 \\
$I_1$ & 240494 & -120117 & 7 \\
$I_2$ & 246363 & -123052 & 7 \\
$I_3$ & 247179 & -123421 & 9 \\
$Q_4$ & 254593 & -127166 & 7 \\
$R_g^2$ & 258044 & -128892 & 7 \\
$a$ & 263911 & -131787 & 9 \\
$N_{\rm{comp}}$ & 266048 & -132894 & 7 \\
\hline \hline
\end{tabular}

\caption{Optimum BIC scores and the corresponding likelihoods, as well as the string length $M$ at which the optimum was achieved, for all the investigated collective variables. Note that smaller BIC values are better.}
\label{tab:likelihoodscores}
\end{table}

In Table~\ref{tab:likelihoodscores}, we have listed the top likelihood scores and the corresponding BIC for the combination of the potential energy with all the other variables described in Sec.\,\ref{subsec:orderparameters}, as well as the string length $M$ at which this value was reached. Note that the minimal BIC is always reached at values of $M > 2$. In other words, in any case, the performance of the optimum string coordinate is better than the corresponding linear reaction coordinate, even after correcting for the higher model complexity introduced by the additional degrees of freedom in the form of the coordinates of the string images as well as the parameters of $f(\alpha)$. Furthermore, as it is the case with the linear version of the reaction coordinate, the combination with the global order parameter $Q_6$ gets the highest score. Even with the added flexibility of the string, no combination of the energy with any other variable than $Q_6$ performs better  than ($U$, $Q_6$) even in the linear case. However, the combinations ($U$,~$N_{\rm{core}}$) and ($U$, $N_{\rm{cryst}}$) come very close, indicating that the flexibility of the string can compensate for the lower quality of the order parameter combinations up to a certain point. Similarly, the likelihood score for the combination with the squared radius of gyration $R_g^2$ is comparatively low even for the best string coordinate.

As an illustration, we have plotted four optimized strings in the $U$-$Q_6$ plane in Fig.\,\ref{fig:string_image}. Shown in the same figure are four optimized strings in the $U$-$N_{\rm{core}}$ plane, the variable combination that gave the second-best likelihood. Note that for larger values of $M$ the strings are distinctly curved, deviating strongly from the linear reaction coordinate studied earlier. Also shown as colormap in the same figure is a comparison of the predicted committor values of the used configurations with the actual one. The mappings $r(\alpha)$ corresponding to the $U$-$Q_6$ strings are shown in Fig.\,\ref{fig:ralpha}.

\begin{figure}
\includegraphics[width=1.0\columnwidth]{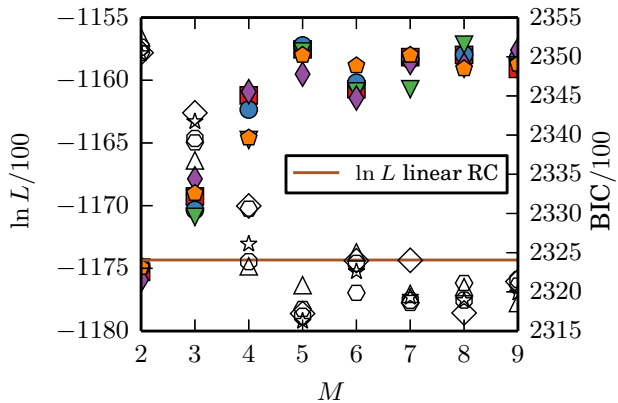}
\caption{Logarithmic likelihood (filled symbols, left scale) and BIC (open symbols, right scale) as a function of string size $M$ for five independent runs of the optimization procedure, with the potential energy $U$ and $Q_6$ as variables. The minimal BIC is reached for ${M=5}$. The likelihood score for the linear model reaction coordinate is also shown for comparison.}
\label{fig:likelihood_M}
\end{figure}

\begin{figure}
\includegraphics[width=1.0\columnwidth]{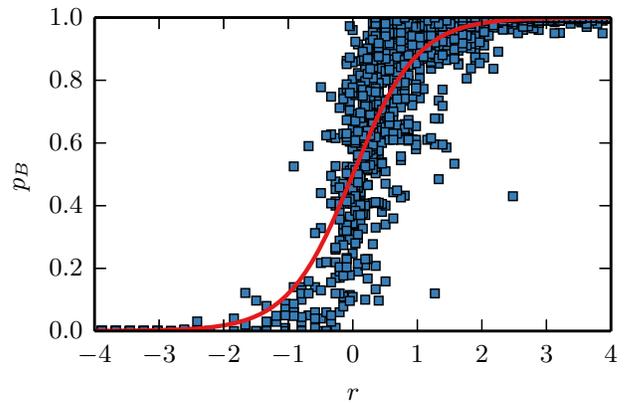}
\caption{The computed committor values plotted against the optimum model reaction coordinate as given by the ${M=5}$ string in the $U$-$Q_6$ plane. Shown in red is the ideal model committor of Eq.\,\eqref{eq:pbofr}.}
\label{fig:pb_of_r}
\end{figure}

The likelihood score, as well as the Bayesian information criterion, as a function of $M$ for a number of independent optimization runs---each with $Q_6$ as second variable---is shown in Fig.\,\ref{fig:likelihood_M}. The ${M=2}$ scores correspond to the linear reaction coordinate. One can observe that the score reaches a plateau for ${M=5}$, in other words, it is sufficient to use a string with that length. Any further addition of string points only increases the model complexity without improving its accuracy, a fact which is also conveyed by the Bayesian information criterion. In Fig.\,\ref{fig:pb_of_r}, we have plotted this optimum reaction coordinate given by the M=5 string---as obtained by one of the optimizations runs---against the real committor value. For a perfect fit, the real committor values would coincide with the hyperbolic tangent function shown in red in the same figure.

\section{Discussion}
\label{sec:discussion}

For sufficiently narrow attractive wells, the polymer chain investigated in this paper shows a two-state folding transition from the expanded coil to the crystalline state. In the present work, using likelihood maximization, we have constructed a string reaction coordinate for this process. As already observed in our previous study using a purely linear reaction coordinate, the combination of the potential energy with the global order parameter $Q_6$ gave the best likelihood score.  Due to the form of the pair potential, the potential energy of the system is basically a measure for the number of contacts between non-neighboring monomers. Moreover, $Q_6$, which is sensitive to closed-packed structures, adds information about the crystallinity of a given configuration. Therefore, it is sensible that a combination of these two parameters will work rather well as a reaction coordinate, since during a typical folding transition, both the number of contacts as well as the crystalline order will increase. Due to the curved nature of the string, our string reaction coordinate is able to follow this transition more closely than a purely linear reaction coordinate. This is what leads to the observed improvement in the likelihood score. In particular, upon following the string along the crystallization, one observes a change of behavior from the initial to the final stages of the transition. Initially, the system changes mainly by increasing its overall crystallinity as quantified by the $Q_6$ parameter. Presumably, this is due to the formation of a small crystalline core in the system. Later on in the crystallization, the strongest change is seen in the potential energy, caused by a steady growth of the initial crystalline core leading to more and more particles packed closely together.

However, the total improvement of the quality of the found reaction coordinate is rather modest. It remains therefore a challenging task to identify better order parameters as candidates in the construction of reaction coordinates. More specifically, these order parameters, while still being as symmetric as possible, should also take into account the order of particles along the polymer chain, which has been completely neglected so far. In the case of more complex polymers, for example if there is less energetic difference between unfolded and crystalline state, it might also help to work with the connectivity information in a more detailed way, rather than with the number of all connections alone.

\begin{acknowledgments}
We thank the Initiativkolleg “Computational Science” of the University of Vienna and the Austrian Science Fund (FWF) within the SFB ViCoM (grant no. F41) and project P 25454-N27 for financial support. The computational results presented have been achieved using the Vienna Scientific Cluster (VSC).
\end{acknowledgments}

\end{document}